# Quasiparticle relaxation dynamics in cuprates and lifetimes of low-energy states: Femtosecond data from underdoped to overdoped YBCO and mercury compounds.

D. Mihailovic[a], J.Demsar[a], R. Hudej [a], V.V.Kabanov [a], T.Wolf[b], J.Karpinski[c]
[a]*Jozef Stefan Institute, Jamova 39, 1001 Ljubljana, Slovenia*
[b]*Forschungszentrum Karlsruhe, Institut fur Technische Physik, D-76021,Karlsruhe,* Germany
[c]*Festkorperphysik, ETH CH-8093 Zurich, Switzerland*

We show that low-energy spectral features in the cuprates can be separated into different components by the measurement of the recombination dynamics of different low-energy excitations in real-time using femtosecond laser spectroscopy. Quasiparticle (QP) recombination across the gap and intra-gap localized state relaxation processes exhibit qualitatively different time- and temperature-dependences. The relaxation measurements also show the existence of two distinct coexisting energy gaps near optimum doping and in the overdoped region, one more or less temperature independent (which exists above and below $T_c$ ) and one which closes at $T_c$ in a mean-field like fashion. Thus systematic studies of QP recombination as a function of doping and temperature suggest that the ground state of the cuprates is a mixed Boson-Fermion system with localised states present over the entire region of the phase diagram.

## 1. INTRODUCTION

From measurements of the time- and temperature- dependence of photoexcited quasiparticle population we can directly deduce the intrinsic QP lifetimes. In addition, a significant amount of detailed information about the low-energy electronic structure can be determined on the basis of studies of QP relaxation as a function of temperature and doping. Although the optical technique for the measurement of QP relaxation is rather new, a number of features can be discerned from the data, which are model-independent and thus interpretation-free. Furthermore, the model for the *T*-dependence of QP relaxation, which has recently been developed has by now been tested on some well-known systems (other than cuprates) exhibiting a low-temperature energy gap[1], so that the results of modeling in conjunction with the data can now be discussed with some confidence.

Here we first give an outline of the universal features of the data in the cuprates, with particular reference to YBCO and Hg1223 and after a brief description of the model used for analysis, we will discuss the implications of the data for our understanding the low-energy electronic structure in these materials.

## 2. QP RELAXATION DATA

Time resolved measurements have been performed in a number of different systems and one can now draw some conclusions about the universality of the observed behaviour. Although the most extensive measurements as a function of doping have so far been performed on O-underdoped[2], optimally doped[3] and Ca-overdoped YBCO[4], temperature dependence measurements have also been performed on BISCO[5], Tl1201[6], Tl2223[7] and Hg1223[8]. Apart from YBCO, in most of these cases the samples investigated were near optimum doping, and so far no systematic doping-dependence data is yet available. However, the photoinduced response (reflectivity or transmittivity) in all of these systems is very similar near optimum doping (see Fig.1). Typically three components with different lifetimes can be discerned. Two are on the femtosecond or picosecond timescale[4] and can be deconvolved only when either their lifetimes are sufficiently different, or they have different sign[6], while the third is substantially longer-lived, on the scale of tens or hundreds of nanoseconds or even more[2,3]. The sign of the two fast components of the photoinduced signal is sometimes different; depending on material and probe wavelength, but the observed temperature dependence of the magnitude of the two fast signals is quite universal. One (with typically $\tau \sim 2$ ps) usually disappears fairly abruptly close to $T_c$, while the other (with typically $\tau \sim 0.4$ ps) diminishes asymptotically up to a temperature $T^*$, which is well above $T_c$ and is dependent on doping[4].

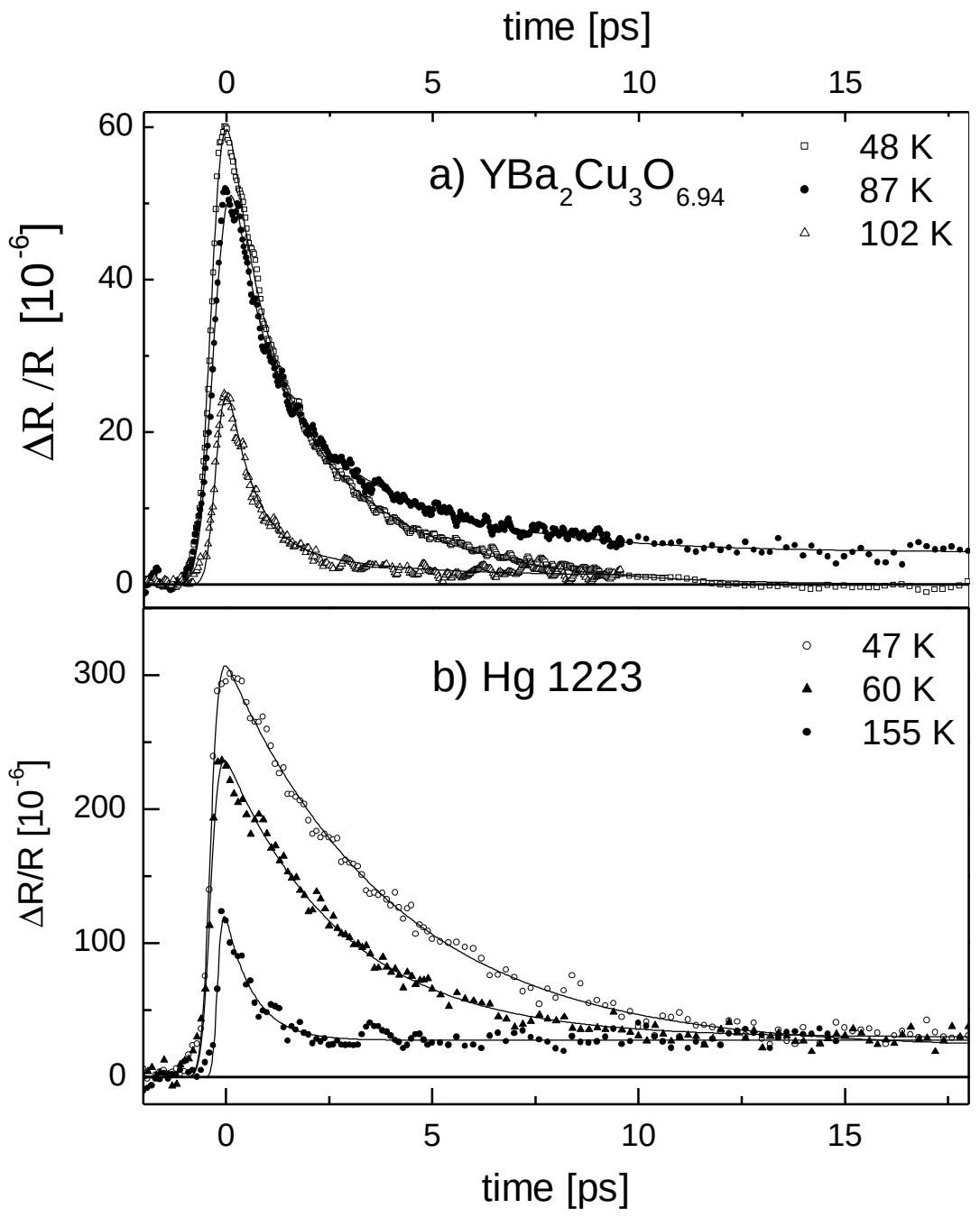

Figure 1. Photoinduced reflection as a function of time in YBCO and Hg1223, below and above $T_c$.

## 3. THE DOUBLE GAP STRUCTURE

The data for photoinduced quasiparticle relaxation have been successfully analysed using a model by Kabanov et al[2] which predicts the temperature dependence of the magnitude of the photoinduced signal for a gap in the density of states, with either a BCS-like T-dependence or a T-independent gap. It also gives an expression for the T-dependence of the relaxation time. The details of the fits to the underdoped and overdoped YBCO data and Hg1223 are presented elsewhere[2,4,8].

The main conclusions reached from the universal features of the picosecond and femtosecond timescale data regarding the low-energy electronic structure are (i) the existence of a $T$-independent gap $\Delta_p$ associated with the subpicosecond relaxation component in the underdoped cuprates which - at least in YBCO, BISCO, Tl1201, Tl2223 and Hg1223 - appears to persist into the optimally doped and overdoped region and is associated with $T^*$. (ii) near optimum doping and in the overdoped region an additional $T$-dependent gap $\Delta_c(T)$ is inferred from the measurements which - on the basis of a divergence of the relaxation time near $T_c$ and the fact that the signal disappears near $T_c$ - is attributed to a *collective* superconducting gap which closes at $T_c$. The two-gap behaviour, which occurs in the optimally doped and overdoped regions, is quite clearly manifested as a two-component QP relaxation in the femtosecond experiments and is quite independent on any model or particular interpretation of the data. We mention here that similar behaviour as observed in the overdoped and optimally doped cuprates, namely a divergence of the QP relaxation time and a disappearance of the signal associated with relaxation across a T-dependent collective gap is observed also in the charge-density wave insulator $K_{0.3}MoO_3$, which is known to exhibit a BCS-like CDW gap[1].

The magnitude of the $T$-independent gap $\Delta_p$ and collective gap $\Delta_c(T=0)$ can be obtained as a function of doping by using the model of Kabanov et al[2]. As seen in Figure 2a), for the case of underdoped YBCO the value of the $T$-independent optical gap (sometimes called a "pseudogap") decreases approximately inversely with the increased doped hole concentration in agreement with a variety of other measurements[9]. In the optimally doped region this $T$-independent gap is nearly constant with doping (within experimental error). The collective superconducting gap $\Delta_c$ in YBCO appears at an oxygen concentration of approximately 6.85 and is present throughout the Ca-overdoped region. It is probably significant that the magnitude of $\Delta_c$ and $\Delta_p$ is very similar in the optimally doped and overdoped regions of the phase diagram (Figs. 2a) and b)). The observed crossover from a single pseudogap to a two-gap structure is thought to reflect a transition from a pre-formed pair scenario in the underdoped state to a collective pairing in a BCS-like scenario at optimum doping and in the overdoped state[10]. Indeed the condensation energy as determined from measurements of the specific heat[11] is large only in the overdoped and optimally doped phases, and drops dramatically in the underdoped phase. The existence of a condensation energy signifies pairing *at* $T_c$, and thus a BCS-like collective pairing scenario. The absence of a condensation energy in the underdoped phase on the other hand is consistent with the pre-formed pair scenario, where no significant change in the DOS occurs at $T_c$ itself.[10]

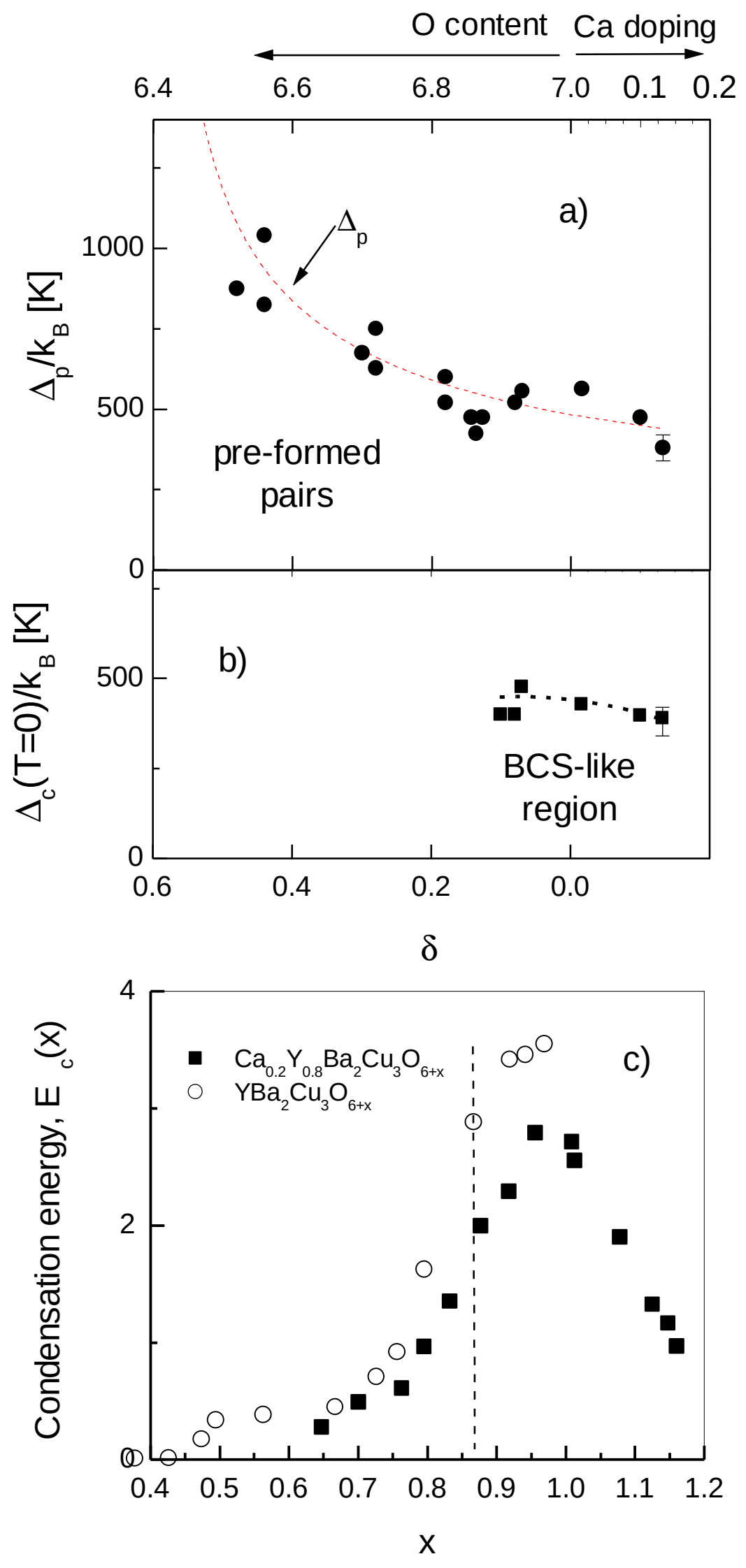

Figure 2. a) The T-independent gap $\Delta_p$ and b) collective gap $\Delta_c$ in YBCO as a function of doping. c) A large condensation energy derived from specific heat by Loram et al[11] is observed only in the region where $\Delta_c$ exists, consistent with the existence of a cross-over to a BCS-like pairing and condensation mechanism near optimum doping.

The observation of a collective gap in the overdoped state (Fig. 2b)) where the condensation energy is large thus strongly supports the proposed picture of the evolution of the low-energy electronic structure[10].

It now remains to be shown by systematic studies of other materials as a function of doping to what extent this behaviour is universal in the cuprates.

## 4. LOCALISED STATES AT OPTIMUM DOPING AND IN THE OVERDOPED REGION

The first indication of the existence of localised states in optimally doped cuprates came from Raman relaxation measurements. Photoexcited carriers were found to display *thermally activated* relaxation in optimally doped YBaCuO rather than temperature-*independent* relaxation expected for intra-band relaxation[12]. The existence of localised states was further confirmed by the presence of slow relaxation processes in the time-resolved photoinduced absorption studies[13,3,2,14]. As already mentioned, the third component evident in the time-domain data has a lifetime which is very long compared to the other two components discussed so far. The amplitude of this component as a function of temperature shows very unusual behaviour, which is strongly doping-dependent. Typically it shows a *peak* on either side of $T_c$, depending on the doping level. The data for overdoped and underdoped CaYBCO are shown in Figure 3. The fits to the data are from the model calculation of Kabanov et al.[14] with two coexisting gaps in the overdoped case (Fig. 3a)) and a single *T*-independent gap in the underdoped case (Fig.3b)). The values of the gaps are the same as those used for fitting the fast QP relaxation data. This gives us confidence that two models are consistent with each other and furthermore, that the analysis of the data is also *quantitatively* correct.

Although one might expect on the basis of its long lifetime that the signal might arise from localised states associated with defects, preliminary experiments on electron-irradiated samples with suppressed $T_c$ have shown a *decrease* in the amplitude of the signal, not an increase, which would expect for a defect-induced signal.[15] This would suggest an intrinsic mechanism for localisation, possibly associated with *d*-wave nodes as suggested by Feenstra et al.[16] or some more intricate localisation mechanism.

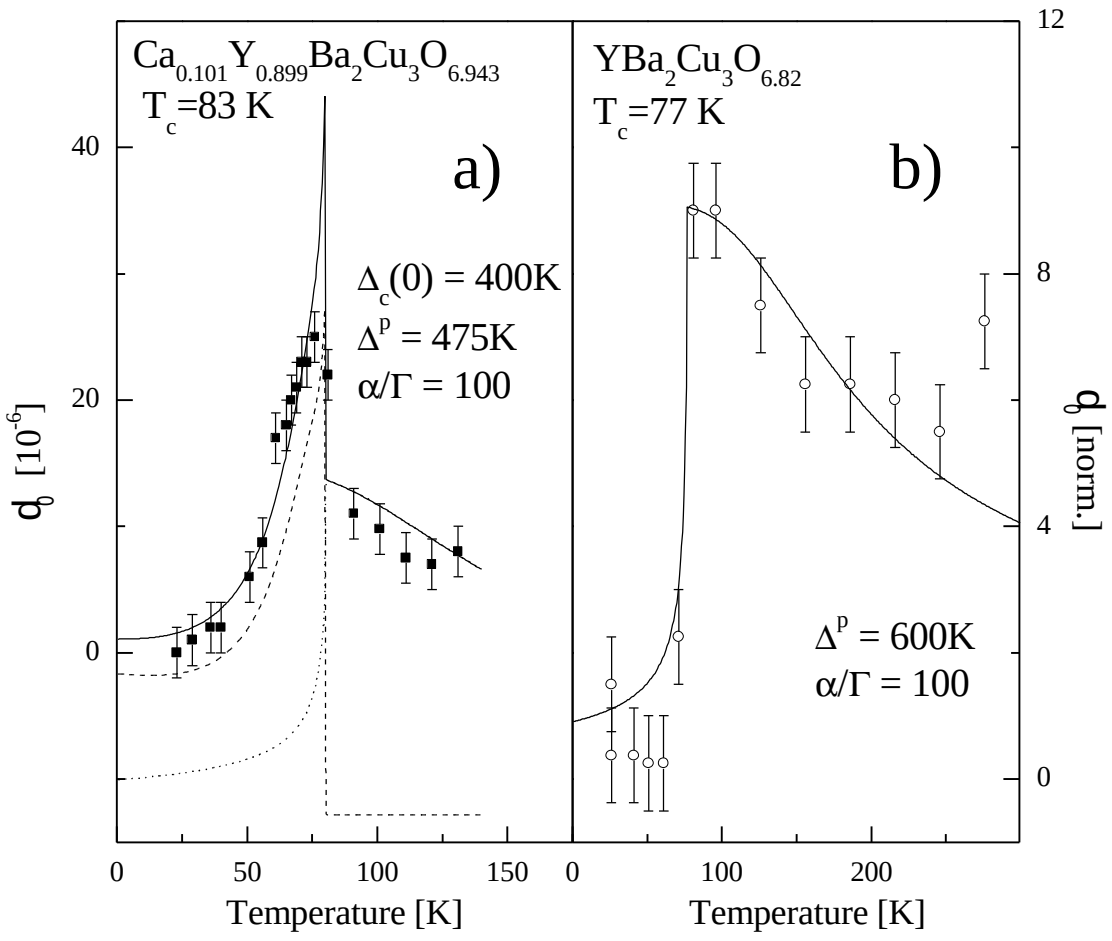

Figure 3. The amplitude of the slow component due to localised states in a) overdoped Ca:YBCO ($T_c$=83 K) and b) underdoped YBCO ($T_c$=77K). The fits are from the model described in ref. 14.

## 5. DISCUSSION

A novel feature of the time-domain investigation of QP dynamics is the existence of a large gap for QPs, which is apparently incompatible with a simple *d*-wave picture of the gap structure[2]. In this case all the QPs are expected to be in the nodes very soon after relaxation (we know that the momentum relaxation is extremely rapid with $\tau_m$ < 10 fs from transport measurements). Clearly the data suggest that the QPs in the cuprates behave in a non-trivial way in the sense that the nodes appear to present a forbidden area of *k*-space for the nonequilibrium QPs. The alternative that the nodes do not exist at all is contrary to many other experimental data showing *d*-wave behaviour, but cannot be dismissed entirely, considering some more recent c-axis tunneling experiments[17].

One possibility is that the time-resolved experiments are sensitive to certain parts of the Fermi surface and not to others. Thus the probe matrix elements or a particular optical resonance could in principle exclude observation of photoexcited QPs in the nodes, while favouring the antinodes. However, the TR measurements have by now been performed at a number of different probe energies ranging from 0.8 eV to 3 eV, and apart from a difference in intensity[3], all the measurements appear to be qualitatively similar. This effectively excludes any role of the probe resonance with the antinodes in relation to the interpretation of the measurements.

Another possibility would be that there is a bottleneck in the intra-node QP relaxation because of the kinematic constraints[16]. The problem with this interpretation is that there would be no sharp gap visible in the experiments, and hence no clear divergence of the relaxation time. Moreover, the existence of well-defined exponential decays is not consistent with this picture. If there is no well-defined gap, the DOS only gradually falls with decreasing energy, so no well-defined decay time is expected.

We thus conclude that the QP relaxation in the cuprates behaves as if the nodes were not present, or are not accessible to the relaxing QPs. Why this happens in not yet entirely clear.